\documentclass{emulateapj}
\usepackage{latexsym}
\usepackage{dcolumn}% Align table columns on decimal point
\usepackage{bm}% bold math
\input{epsf}
\newcommand{\be}{\begin{equation}}
\newcommand{\ee}{\end{equation}}
\newcommand{\ba}{\begin{eqnarray}}
\newcommand{\ea}{\end{eqnarray}}
\newcommand{\bfi}{\begin{figure}
\epsfxsize=9cm
\epsffile}
\newcommand{\efi}{\end{figure}}
\newcommand{\apjl}{ApJ}
\newcommand{\apjs}{ApJS}
\newcommand{\aj}{AJ}
\newcommand{\mnras}{MNRAS}

\begin{document}
\title{Self calibration of galaxy bias in spectroscopic redshift
  surveys of baryon acoustic oscillations}   
\author{Pengjie Zhang}
\affil{Shanghai Astronomical Observatory, Chinese Academy of
  Science, 80 Nandan Road, Shanghai, China, 200030}
\affil{Joint Institute for Galaxy and Cosmology (JOINGC) of
SHAO and USTC}
\begin{abstract}
Baryon acoustic oscillation (BAO) is a powerful probe on the expansion
of the  
universe, shedding light on  elusive dark energy and gravity at
cosmological 
scales. BAO  measurements through biased tracers of the underlying
matter density field, as most proposals do,  can reach high statistical
accuracy. However,  possible scale dependence in bias may induce
non-negligible systematical errors, especially for
the most ambitious spectroscopic surveys proposed. We show that
precision spectroscopic redshift information available in 
these surveys allows for {\it self calibration} of the galaxy bias and its
stochasticity, as function of scale and redshift. Through the effect
of redshift  distortion,  one can simultaneously measure the real
space power 
spectra of galaxies, galaxy-velocity and velocity, respectively. At
relevant scales of BAO, galaxy velocity faithfully traces that of the
underlying matter. This valuable feature enables a rather model
independent way to measure the  galaxy bias and its stochasticity  by
comparing the three power spectra. For the square kilometer array
(SKA), this   self calibration is statistically accurate to correct for $1\%$ 
level shift in BAO  peak positions induced by bias scale
dependence. Furthermore, we find that SKA is able to detect BAO in the
velocity power spectrum, opening a new window for BAO cosmology. 
\end{abstract}
\maketitle
\section{Introduction}
The universe is  the  largest laboratory for fundamental physics. The zeroth order 
descriptions of the universe include the overall expansion rate $H(z)$ and its
integral form, the distance as a function of redshift $z$.  Type Ia
supernovae (SNe Ia), as  promising {\it cosmological 
standard candles}, allow  the measurements of cosmological distance and
have enabled the discovery of the late time acceleration of the
universe \citep{Riess98,Perlmutter99}.  This discovery has
profound implication for fundamental physics, leading to either a
dominant dark 
energy component with equation of state 
$w\equiv P/\rho<-1/3$ or significant deviations from general
relativity around 
Hubble scale. Ongoing and planned supernova surveys will be able to
significantly narrow the range of  $w$ and hopefully clarify the role of the 
cosmological constant in our universe. Meanwhile, other independent probes are
indispensable, to 
reduce statistical errors, break parameter degeneracies and to cross check for
possible systematical errors.  The baryon acoustic oscillation (BAO) is
such a probe \citep{Blake03,Hu03,Seo03,Seo05,Blake06}. 

Prior to the epoch of recombination, baryons and photons are tightly
coupled.  Fluctuations in the baryon-photon fluid
propagate as acoustic waves at speed comparable to  the speed of
light. These acoustic waves imprint in the cosmic microwave
background, a snapshot of the  photon fluid at the
epoch  of decoupling. They also imprint in the baryon fluid through 
coupling between photons and baryons and later in the overall matter
density field, through gravitational coupling between baryons and dark
matter particles \citep{Hu96,Eisenstein98}.  These BAO features have
been detected and measured to high precision in CMB and have put
strong constraints on the  geometry of the universe
\citep{WMAP3}. In the large
scale structure (LSS) of the universe, these features are
significantly suppressed by dark matter  to a level much weaker than
that in CMB. Nonetheless,  they are detectable, as shown in
measurements of correlation function and power spectrum of 2dF and SDSS galaxies
\citep{Cole05,Eisenstein05,Percival07}. The comoving scales
of BAOs can be predicted accurately from well understood linear physics. Thus we
are able to convert the measured angular separations of these features
on the sky to the cosmological angular diameter distance, free of many
astrophysical uncertainties.  Furthermore, 
since we can measure their separations in 
redshift, we are able to directly measure $H(z)$, which is more
sensitive to $w$ than the distance. For these merits, BAO is 
widely accepted as a robust and powerful {\it cosmological standard ruler} to
reveal the expansion  history of the  universe \citep{DETF}.

 All existing proposals on BAO measurements target at density field of
 various tracers of the LSS. Most are galaxy spectroscopic redshift
 surveys, including 
LAMOST\footnote{http://www.lamost.org/en/}, WiggleZ
\citep{Glazebrook07},
BOSS\footnote{http://cosmology.lbl.gov/BOSS/},  WFMOS \citep{Bassett05},
HETDEX\footnote{http://www.as.utexas.edu/hetdex/},
ADEPT\footnote{http://www7.nationalacademies.org/ssb/BE\_Nov\_2006\_bennett.pdf}
and the square kilometer array (SKA, \citealt{Abdalla05}). Besides, there are
proposals to measure BAO in the 21cm background, both at low
redshifts \citep{Chang07} and at high redshifts
\citep{Mao07,Wyithe07}, in the Lyman-alpha forest \citep{McDonald07}
and in SNe Ia \citep{Zhan08}. Sub-percent level statistical 
accuracy can be reached in the most ambitious surveys, such as
SKA, classified as a stage IV project by the dark energy task force
\citep{DETF}. However, possible 
sources of systematical errors could prohibit us to exploit full
advantage of these measurements. Some, for example those induced by 
gravitational lensing \citep{Vallinotto07,Hui07a,Hui07b,LoVerde07}, are
correctable \citep{Zhang07b,Hui07b}.  Nonlinear evolution in the matter
density field is also  likely correctable, either by improvement in
theoretical calculation
\citep{Crocce06,Crocce07,Matarrese07a,Matarrese07b,McDonald07a} or in
data analysis \citep{Eisenstein07b}.  However, the clustering bias of
these tracers  with respect to the underlying matter density field, is
more elusive to handle.  The scale
dependence of the bias changes the overall shape of the power
spectrum and thus shifts the BAO positions. A good thing is that,  to
the first order accuracy, BAOs, as periodic sharp features on top of  the 
smooth power spectrum, are robust against bias scale dependence, which
is likely more smooth.  For this reason and others, systematical errors
caused by possible scale dependent 
bias are likely sub-dominant, comparing to the statistical errors, for
BAO measurements of existing surveys and  proposed medium
cost surveys.  On the other hand, the most ambitious surveys such as
SKA  will have about two orders of magnitude  increase in survey volume and
one order of magnitude increase in redshift over SDSS. For these
surveys, statistical errors in the distance and 
$H(z)$ measurement will be reduced to below $1\%$ \citep{Blake06}. At
this stage,  shifts in BAO peaks induced by even
rather mild scale dependence in galaxy bias may become non-negligible. For discussions on the effect of bias to BAO, refer to
e.g. \citet{Seo05,Eisenstein07a,Guzik07,Huff07,Smith07a,Angulo08}. 

Fortunately, spectroscopic redshift information, which is required to
measure $H(z)$ along the line of sight through BAO, allows for self
calibration of galaxy bias through the effect of redshift distortion.
Spectroscopic redshift surveys are able to  measure $P_g^s({\bf k})$, the 3D
power spectrum of galaxies in redshift  space, to 
high precision.  Galaxy peculiar velocities make $P_g^s$ anisotropic
in the 3D wavevector ${\bf k}$ space. This redshift distortion effect
is often treated as a source of error. However, it can be rendered
into valuable source of signal instead, given sufficiently accurate
understanding of redshift distortion.  

 Through the anisotropy in redshift space, one can reconstruct the
 real space velocity power spectrum 
$P_{\theta}$, the velocity-galaxy cross correlation power spectrum
$P_{g\theta}$  and the galaxy power spectrum $P_g$ in a rather model
 independent way at cosmological distances. Here $\theta$ is the normalized
 divergence of peculiar velocity. Since 
galaxies have rather low mass, they  are sub-dominant in gravitational
field at scales $\ga 10 $Mpc.  At scales of BAO (much larger than $10$ Mpc,
even for the third peak), gravity is
the only force accelerating  galaxies.  For these reasons, galaxies
can be well approximated as test particles, following 
the motion of dark matter particles. Thus the large scale galaxy
velocity field is a faithful tracer of the overall matter velocity
field. $P_{\theta}$ and its relation with the matter power 
spectrum $P_m$ can then be predicted  from first principles, free of
uncertainties in understanding of galaxy formation.
Comparing $P_g$, $P_{g\theta}$ and $P_{\theta}$, we are able to
measure galaxy bias and its stochasticity. This allows for self
calibration of galaxy bias in BAO cosmology. In
a pioneer work, \citet{Pen98} proposed to  measure galaxy bias and its
stochasticity through the quadrupole and octupole of  $P_g^s$. This
work extends his by a full power spectrum reconstruction, for which we
are able to utilize information contained in all moments. 

 On the other hand,  BAOs also exist in the
velocity field. $P_{\theta}$ reconstruction thus opens a new window of
BAO measurement. This approach is free of problems of galaxy bias,
although it may not be able to reach the same statistical accuracy as
that of the density field. 

This technique can be  incorporated into existing
BAO analysis methods, such as that of \citet{Seo03,Seo07} and
\citet{Blake03,Blake06,Percival07}, as straightforward post processing. We will
explain this issue later. 

SKA is able to detect billions of
galaxies over  half  sky to redshift  $z\ga 3$ through
neutral hydrogen 21cm emission of these galaxies and measure precise
spectroscopic redshifts without extra cost. For this reason, we choose
SKA as our primary target for investigation. We present the
reconstruction procedure in \S \ref{sec:method},  demonstrate  the feasibility
of SKA to self calibrate galaxy bias in \S \ref{sec:calibration}, 
show that BAOs in the galaxy velocity field are detectable through SKA
in \S \ref{sec:velocity} and discuss in \S \ref{sec:discussion}. 
%%%%%%%%%%%%%%%%%%%%%%%%%%%%%%%%%%%%%%%%
\section{Velocity reconstruction through redshift distortion}
\label{sec:method}
Peculiar motions of galaxies imprint unique signatures in the
redshift space galaxy power spectrum $P_g^s({\bf k})$, which takes a general
form (e.g. \citet{Scoccimarro04}), 
\ba
\label{eqn:RD}
P^s_g({\bf
  k})=\left[P_g(k)+2u^2 P_{g\theta}(k)+u^4
P_{\theta}(k)\right]F\left(\frac{k^2u^2 \sigma^2_v}{H^2(z)}\right).
\ea
Here, $P_g$, $P_{g\theta}$, $P_{\theta}$ are
  the real space power  spectra of galaxies, galaxy-velocity and
  velocity. $\theta$ is the comoving peculiar velocity divergence
divided by $-H(z)$.   $\sigma_v$ is the 1D velocity dispersion; and
$F(x)$ is a smoothing function, normalized to unity at $x=0$,
determined by the velocity probability distribution. $u=k_{\parallel}/k$ is the
  cosine of the angle of the ${\bf k}$ vector with  respect to radial
  direction;  This unique directional dependence has enabled 
  successful simultaneous reconstruction of the three power spectra
 from 2dF and SDSS galaxies \citep{Tegmark02,Tegmark04}.  We adopt a simple
  minimum variance estimator, developed in \citet{Zhang07}, to quantify
  accuracies of this approach.

   For each ${\bf k}_i$ in the given $k$ bin, we have a measurement of
   $P^s_g$, which we denote as $P_i$.  The unbiased minimum variance
  estimator of the band power of $P_{(\alpha)}$
  ($\alpha=g,g\theta,\theta$),  is   $\hat{P}=\sum W^{(\alpha)}_iP_i$, where 
  $W^{(\alpha)}_i=\frac{F_i}{2\sigma^2_i}(\lambda^{(\alpha)}_1+\lambda^{(\alpha)}_2
  u_i^2+\lambda^{(\alpha)}_3 
  u_i^4)$ and  $F_i\equiv F(ku_i\sigma_v/H)$. $\sigma_i=P_i+1/n_g$
   is the rms fluctuation of $P_i$, where $n_g$ is the galaxy number
   density. The Lagrange multipliers $\lambda^{(\alpha)}_{1,2,3}$ are given by
   ${\bf \lambda}^{(g)}=(1,0,0)\cdot {\bf  
   A}^{-1}$, ${\bf \lambda}^{(g\theta)}=(0,1/2,0)\cdot {\bf
   A}^{-1}$  and ${\bf \lambda}^{(\theta)}=(0,0,1)\cdot {\bf
   A}^{-1}$. $\lambda^{(\alpha)}$ are orthogonal to each
   other. The $3\times 3$ matrix ${\bf A}$ is given by
\ba
\ A_{mn}=\sum_i
   u_i^{2(m+n-2)}\frac{F_i^2}{2\sigma^2_i}\ \ ;\ m,n=1,2,3\ .
\ea
The corresponding error in each power spectrum is 
\be
\sigma^{(\alpha)}_p=\left(\frac{1}{2}  \lambda^{(\alpha)}\cdot {\bf A}\cdot
   [\lambda^{(\alpha)}]^{T} \right)^{1/2}\ .
\ee
One can check that it has the right scaling that $\sigma_p\propto
\sigma/\sqrt{N}$, where $N$ is the number of independent modes and
$\sigma$ is some average of $\sigma_i$. This relation can be further
simplified and we have $\sigma_p^{(g)}=\sqrt{\lambda_1^{(g)}/2}$,
$\sigma_p^{(g\theta)}=\sqrt{\lambda_2^{(g\theta)}/4}$ and
$\sigma_p^{(\theta)}=\sqrt{\lambda_3^{(\theta)}/2}$. 

Combining these power spectrum measurements, we are able to measure
the galaxy bias and its stochasticity through
\ba
\label{eqn:calibration}
b_g^2=\left(\frac{P_g}{P_{\theta}}\right)\left(\frac{P_\theta}{P_m}\right)_{theory}\ ;\ r=\frac{P_{g\theta}}{\sqrt{P_gP_{\theta}}}\  \ .
\ea
Here, $(P_\theta/P_m)_{theory}$ is predicted from theory. In
linear regime, this value is $f^{-2}$, where $f\equiv (d\ln D/d\ln
a)^2$ and $D$ is the linear density growth factor. $r$ is the cross
correlation coefficient, whose deviation from unity is a measure of
stochasticity \citep{Pen98,Dekel99}.

This reconstruction requires the angular diameter
distance $D_A$ and $H(z)$ as input to convert the
observed angular and radial separation into ${\bf k}$. For numerical results
presented in this paper, we just take $D_A$ and $H(z)$ of the
fiducial cosmology as input. For real data, a convenient
and self consistent procedure to carry out this reconstruction is as
follows. First,  one can measure the  
distance and $H(z)$  by the usual methods \citep{Blake03,Seo03}. 
 The measured distance
and $H(z)$ are then used as input for the reconstruction of the power
spectra,  which  are in turn  applied to check for the consistency of
scale independent bias. This can be done iteratively. Alternatively,
one can allow for scale dependent bias and its stochasticity in BAO analysis
method of \citet{Seo03,Seo07}. Through simultaneous multi-parameter
fitting, galaxy bias (amplitude, scale dependence and stochasticity)
is automatically self calibrated. 

Errors in $D_A$ and $H(z)$ propagate into
the power spectrum reconstruction, through two effects: a constant
fractional shift in $k$ and a wrong determination in the ${\bf k}$ direction
$u$.  Obviously, the overall shift in $k$ does not affect the
measurement of $b_g$ and $r$.  However, a wrong determination in $u$
does. $P_{g\theta}$ and $P_{\theta}$ are determined by  the derivatives of
$P^s_g(u)$ with respect to $u$.  We can show that the induced errors in the
power spectra have only weak dependence on $k$ through the slope of
the power spectra. Thus the main effect of errors in $D_A$ and $H(z)$
is to bias the overall amplitude of $b_g$ and $r$, instead of
introducing new scale  dependence.  For these reasons,  
it suffices to neglect this kind of error source.

An implicit simplification adopted in this reconstruction is that
$\sigma_v$ is given. In reality, at BAO relevant scales, the function
$F(ku\sigma_v/H)\simeq 1$. Thus our simplification is justified
here. For real data, to improve the accuracy of reconstruction, one
should treat $\sigma_v$ as a free parameter to be marginalized over. We
do not expect any major effect 
on the forecasted errors in the reconstructed power spectra. 

Hereafter we will adopt the linear theory to carry out numerical
calculations. For real data,  nonlinearity must be taken into
account in BAO analysis.  For example, nonlinear evolution in density
field and velocity field are not the same
\citep{Bernardeau02,Smith07b}. As a result, 
$(P_\theta/P_g)_{theory}$ is no longer equal to $f^{-2}$ and is no
longer scale independent. If unaccounted, we may obtain a
false scale dependence in $b_g$. However, this should not
be a real problem, due to three reasons. (1) The
reconstruction does not rely on assumptions of linearity. (2) The bias
is  defined with respect to the real (thus nonlinear) power
spectra. (3) Nonlinear evolution in the 
velocity field and matter field can be calculated from first
principles. As long as we  replace the linear theory version $f^{-2}$
with the actual (nonlinear) $(P_\theta/P_m)_{theory}$ in  real data
analysis, nonlinearity does not bias the result.

\section{Self calibration of galaxy bias}
\label{sec:calibration}
In this section, we choose SKA as the primary target for
investigation. 
For the redshift range $1\la
z\la 2$ of most interest, the SKA survey speed $(A_{\rm eff}/T_{\rm
  sys})^2\times$FOV$=2\times 
10^{10}$ m$^2$ K$^{-2}$
deg$^2$.\footnote{http://www.skatelescope.org/pages/Preliminary\_specifications\_v2.4.pdf. This
survey speed is much higher than what adopted in preivous estimations of
\citet{Abdalla05,Zhang06}, resulting in much higher sensitivity to
detect high redshift galaxies.} 
With this specification, SKA is able to detect several billions of
21cm emitting galaxies in
10  year survey.  Throughout this paper, we adopt the galaxy surface number
density as  $20$ galaxies per square arc-minute, corresponding to 2.4 billion
over three quarter of sky. The exact number is hard to estimate, due to
poor knowledge on these galaxies at high redshifts. A good thing is
that the exact number is not 
necessary for the estimation presented in this paper, due to the
dominance of cosmic variance over shot noise, especially at 
scales larger than the second baryon acoustic peaks. 

To proceed, we adopt the fiducial cosmology as  a flat $\Lambda$CDM
cosmology, with $\Omega_m=0.268,
\Omega_b=0.044,\Omega_{\Lambda}=1-\Omega_m$, $h=0.72$ and
$\sigma_8=0.77$ \citep{WMAP3}. We adopt the transfer function fitting
formula of \citet{Eisenstein98}. We stick to the linear
perturbation theory and neglect 
the nonlinear evolution in the matter density field, since it is predictable
and correctable. The velocity power spectrum is fixed 
through  $P_\theta=f^2P_m$. Here, $f=d\ln D/d\ln a$, $D$ is the linear
density  growth factor and $P_m$ is the matter power spectrum. 

To demonstrate the feasibility of self calibration, we introduce scale
dependence and stochasticity in galaxy distribution. As a reminder,
the galaxy bias $b_g(k,z)$ is defined by  $b^2_g(k,z)\equiv
P_g(k,z)/P_m(k,z)$.   The stochasticity of galaxy distribution is
characterized by the cross correlation coefficient $r(k,z)$, defined
through $r^2\equiv P_{g\theta}^2/(P_gP_\theta)$. Deterministic bias
will have $r=1$. This is likely true at very large
scales. Proceeding to smaller scales we may expect $r<1$, because
galaxy formation depends on not only the local density, but also the
environment. We adopt simple forms of 
parametrization:  $b_g=b_{g0}+db_g/dk|_{k=0}k\equiv b_{g0}(1-k/k_b)$ and
$r=1+dr/dk|_{k=0}k\equiv 1-k/k_c$.  Here, $b_{g0}$ is the galaxy bias
when $k\rightarrow 0$. We further adopt $k_b=4h/$Mpc and $k_r=10
h/$Mpc. The actual scale dependence of bias can be much more 
complicated, with terms $\propto k$, $\propto k^2$, $\propto
1/P_m(k)$, or even more complicated terms (refer to, e.g. 
\citealt{Smith07a}). We
choose the above simple forms of parametrization and values for several
reasons. (1) These forms of parametrization are just the Taylor
expansion of $b_g$ and $r$ at $k=0$ to first order. Despite their
simplicity, they are rather general to describe small deviations from
scale independence. (2) For
the values adopted, the galaxy bias decreases by less than $5\%$
over the first three BAO peaks. This scale dependence has the same
sign as, but smaller amplitude than, what predicted for low mass
galaxies \citep{Sheth99,Guzik07,Smith07a}. Thus the
assumed scale dependence likely represents a conservative lower limit
of this source of systematical errors. (3) Such level of
scale dependence in $b_g$ may cause non-negligible systematical
errors. It shifts the positions of the second and the third peak by
$\sim 1\%$. This level of systematic shift in BAO positions could overwhelm the
statistical uncertainties in SKA BAO measurements.\footnote{The exact
  level of systematical errors induced by scale dependent bias varies
  with BAO analysis methods. The method of \citet{Seo03} performs the
  global fitting to the galaxy power spectrum and  utilizes the
  cosmological  information in the overall shape of the power
  spectrum. On the other  hand, the method of
  \citet{Blake03,Percival07} discards 
  such information. It fits the overall shape of $P_g$ with a smooth
  reference function $P_{\rm ref}$. It then fits $P_g/P_{\rm ref}$ by
  a decaying sinusoidal functiion, in which the BAO information is
  encoded. $P_{\rm ref}$ can  capture and thus filter away some of, if
  not most of,  the  bias scale dependence. Thus 
  this method is less affected by the scale  dependence in galaxy
  bias.   But the residual could still bias the distance measurement. } For 
these reasons, the   assumed bias scale dependence  and stochasticity
are suitable to demonstrate the feasibility of the self calibration technique. 

We remind the readers that, these parametrizations of $b_g(k,z)$
and $r(k,z)$ only serve as input of  the fiducial model.   The reconstruction of
$P_g$, $P_{g\theta}$ and $P_\theta$ does not rely on any assumptions on $b_g$
and $r$. Parametrizations on $b_g$ and $r$ can  certainly improve over
the statistical  errors in the reconstruction. However, inappropriate
parametrizations could bias the reconstruction. For this reason,  we
do not explore such approach in this paper.

%%%%%%%%%%%%%%%%%%%%%%%%%%%%%%%%%%%%%%%%%%%
\bfi{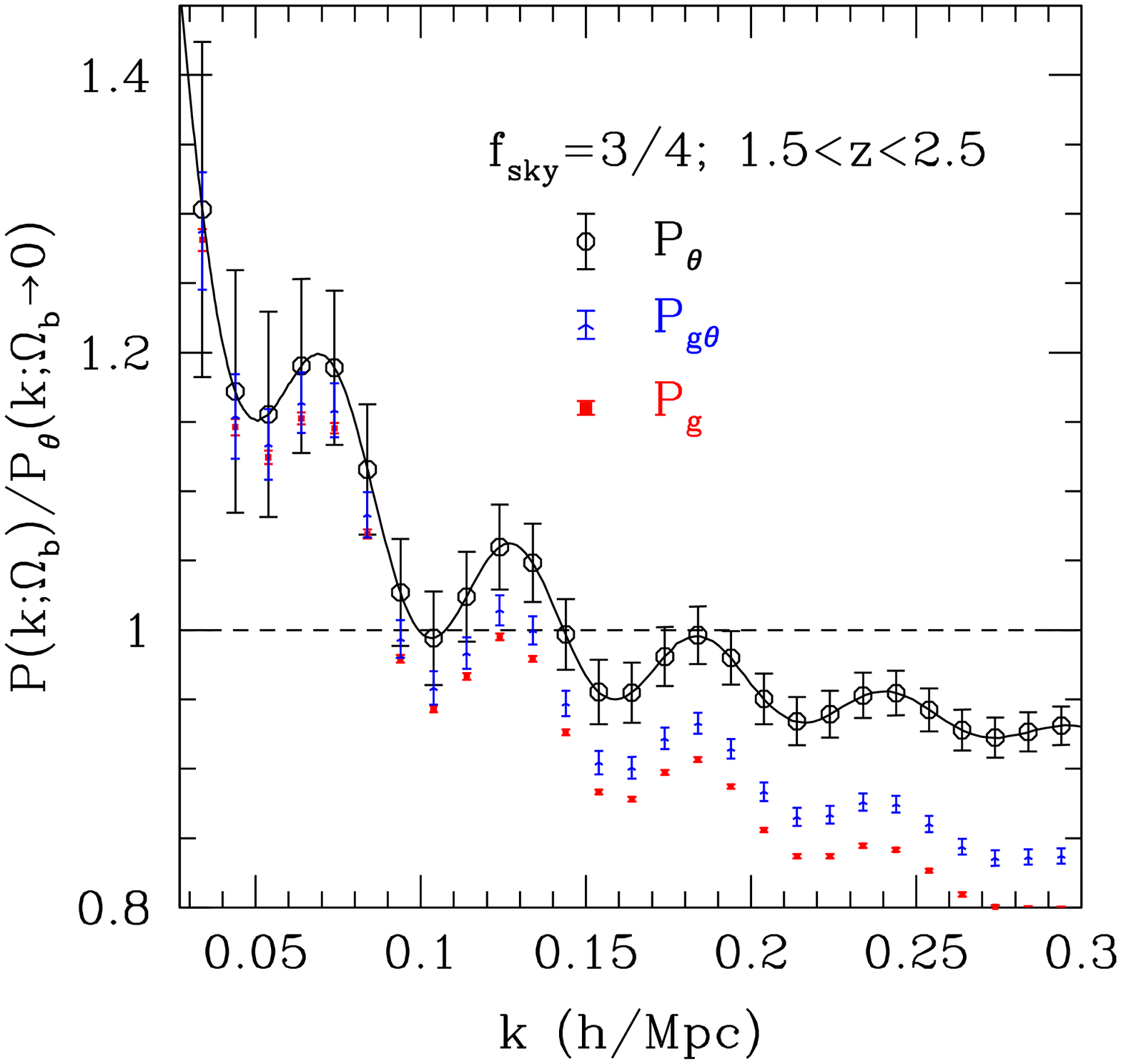}
\caption{Forecasted accuracies in reconstructed power spectrum $P_g$,
  $P_{g\theta}$ and $P_{\theta}$. The estimation is based on SKA ten 
  year survey over three quarters of sky and galaxies at $1.5<z<2.5$. We plot the ratio of three power
  spectra with respect to $P_{\theta}(\Omega_b\rightarrow 0)$ and
  normalized the ratios to be equal at $k\rightarrow 0$. Error bars 
  are forecasted adopting $b_{g0}=1$.  $P_{\theta}$ is much harder to
  measure than $P_g$ and $P_{g\theta}$, so the self calibration is
  fundamentally limited by the associated error in
  $P_{\theta}$. Nonetheless, SKA is able to test and
  correct for several percent level scale dependence in bias across
  the first three peaks.  Furthermore, with such accuracy, BAOs are
  detectable in $P_\theta$. Accuracy in $P_{\theta}$ is sensitive  
  to $b_g$, as shown in Fig. \ref{fig:v1}. \label{fig:v2}}
\efi
%%%%%%%%%%%%%%%%%%%%%%%%%%%%%

For SKA,  $P_g$  can be reconstructed to high precision
(Fig. \ref{fig:v2}) and the associated fractional  statistical errors are very
close to that of corresponding $P_g^s$. Accuracies in $P_{g\theta}$
reconstruction are about a factor of $5$ poorer, but still
impressive. A further factor of $\sim 3$ degradation occurs in $P_{\theta}$
reconstruction. This is what we expected. The average contribution of
$P_{\theta}$ to $P_g^s$ is $P_{\theta}\langle
u^4\rangle=P_{\theta}/5\simeq P_m\Omega_m^{1.2}(z)/5$ and that of
$P_{g\theta}$ is $2P_{g\theta}\langle u^2\rangle\simeq P_m
2\Omega_m^{0.6}(z)/3$. The realistic galaxy bias is likely $b_g\ga
1/2$ (Fig. \ref{fig:bias} and further reading in \citealt{Jing98}). We
thus have $P_{\theta}\langle u^4\rangle <2P_{g\theta}\langle 
u^2\rangle<P_g$. Furthermore, we rely on the
directional dependence of $P_g^s$ to  extract sub-dominant
components $P_{g\theta}$ and
$P_{\theta}$, resulting in $\sigma_p^{(\theta)}/P_{\theta}>\sigma_p^{(g\theta)}/P_{g\theta}>\sigma_p^{(g)}/P_{g}$.

Nonetheless, the reconstructed power spectra (especially
$P_{\theta}$) have the statistical accuracy to detect $5\%$ or even smaller
variation in the  galaxy bias over the first three peaks, with large
statistical significance (Fig. \ref{fig:v2}). This $5\%$ scale
dependence causes $1\%$ shift in BAO peak positions. Thus the self
calibration technique is able to  reduce systematical error in BAO
peak positions induced by scale dependent bias to smaller than $1\%$.

However, despite its promising capability, the  self calibration
procedure is fundamentally limited by  accuracy in our understanding
of redshift distortion. Systematic deviation in the modelled  $P_g^s$ from the actual
one can propagate into the estimation of $b_g$ and thus can bias the self
calibration of galaxy bias.  A good news is that, to 
measure the scale dependence and perform self calibration, we are not
necessarily confined to BAO scales. As 
seen from Fig. \ref{fig:v2}, we are able to reconstruct the three
power spectra to $k\sim 0.3 h/$Mpc and even smaller
scales. $k=0.3h/$Mpc is still in the linear regime at $z=2$. This means
that we have a factor of $\sim 2$ larger (and still safe) dynamical
range to measure the scale dependence in $b_g$. Furthermore, variation
in $b_g$ over this larger dynamical range is likely larger. Both alleviate the
accuracy requirement for modeling the redshift distortion (e.g. Eq. \ref{eqn:RD}). 
For the simple model we adopted, this means to measure $8\%$ variation
in $b_g$ over $k\la 0.3 h/$Mpc instead of measure $\sim 5\%$ variation
over $k<0.2h/$Mpc. Still,  the modeling
of the redshift space power spectrum must be accurate to much better
than  $2\times 8\%=16\%$ percent in order to render the associated
systematical error negligible. This requirement is  challenging, but not
impossible, given the rapid development  in  simulations and analytical
calculations.  We refer the readers to works on modeling of redshift
distortion, through perturbation theory 
\citep{Scoccimarro04,Matsubara07}, halo model
\citep{White01,Seljak01,Tinker07} 
and N-body simulations \citep{Kang02,Tinker06}.  

%%%%%%%%%%%%%%%%%%%%%%%%%%%%%%%
\bfi{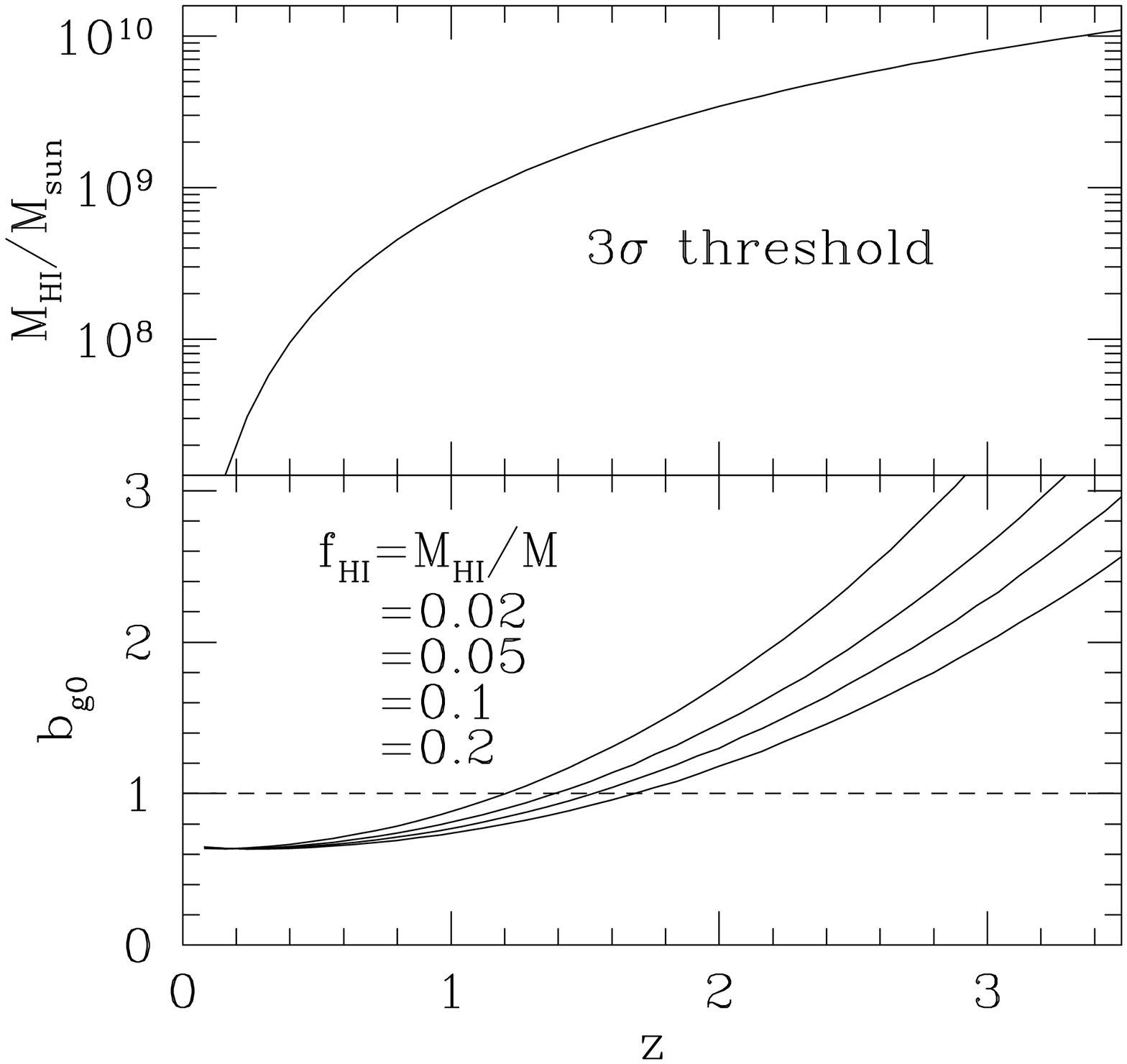}
\caption{The SKA galaxy selection threshold (top panel) and the corresponding
  bias (bottom panel). SKA galaxies are selected by the 
  galaxy  neutral hydrogen mass $M_{\rm HI}$, while the bias is
  determined by the total galaxy mass $M$. Lack of robust measurements on
  $f_{\rm HI}\equiv M_{\rm HI}/M$, we plot the galaxy bias for several
  possible values of $f_{\rm HI}$. \label{fig:bias}}
\efi
%%%%%%%%%%%%%%%%%%%%%%%%%%

\section{Detecting BAO in the velocity power spectrum}
\label{sec:velocity}
BAOs in the density field inevitably show up in the gravitational potential
field, as directly seen from the Poisson equation. This featured gravitational
field accelerates galaxies and thus imprints  BAOs in the velocity
power spectrum $P_{\theta}$.  At scales of BAOs, the galaxy velocity
filed should be unbiased with respect to the velocity field of the
underlying matter, since these galaxies are test particles responding
only to gravity at these scales. For this reason, we are able to
calculate $P_{\theta}$ from first principles. Since in principle there is no
ambiguity in precision calculation of  $P_{\theta}$,  BAO cosmology
based on $P_{\theta}$ is free of the issue of galaxy bias. It is thus
worthy of detecting BAOs in  $P_{\theta}$.  

In reality, BAO measurements through  reconstructed $P_{\theta}$ from redshift
distortion are challenging, since accuracies in $P_{\theta}$ are much
worse than that in $P_g^s$ or $P_g$, as explained in last
section. For $b_{g0}=1$, the fractional error in
$P_{\theta}$ is a factor of $\sim 10$ higher than that in $P_g$
(Fig. \ref{fig:v2}). It thus requires $\sim 100$ times increase in survey volume
in order to reach the same statistical accuracy obtained from the
density field.  
For this reason, BAO can not be detected in $P_{\theta}$ by those low
redshift spectroscopic surveys such as LAMOST and BOSS. 
But SKA can (Fig. \ref{fig:v2} \& 
\ref{fig:v1}), so as the Hubble sphere hydrogen survey (HSHS)
\citep{Peterson06}  and  ADEPT. For SKA,  the statistical accuracy in 
$P_{\theta}$ is about $3\%$ around the second and the third peaks (for
bin size $\Delta k=0.01 h/$Mpc),
slightly larger than that of the $P_g^s$ measurement in the combined
sample of SDSS main galaxies and  luminous red galaxies
\citep{Percival07}.  SKA extends much larger redshift
range than SDSS. Higher redshift means more ${\bf k}$ modes suitable
for BAO detection (less severe smear of BAO features by mode-coupling
in nonlinear regime) and higher sensitivity to dark matter and dark
energy. Thus  BAO cosmology through the recovered $P_{\theta}$  may gain over
current SDSS BAO measurement through $P_g^s$.  

%%%%%%%%%%%%%%%%%%%%%%%%%%%%%%%%%%%%%%%%%%%
\bfi{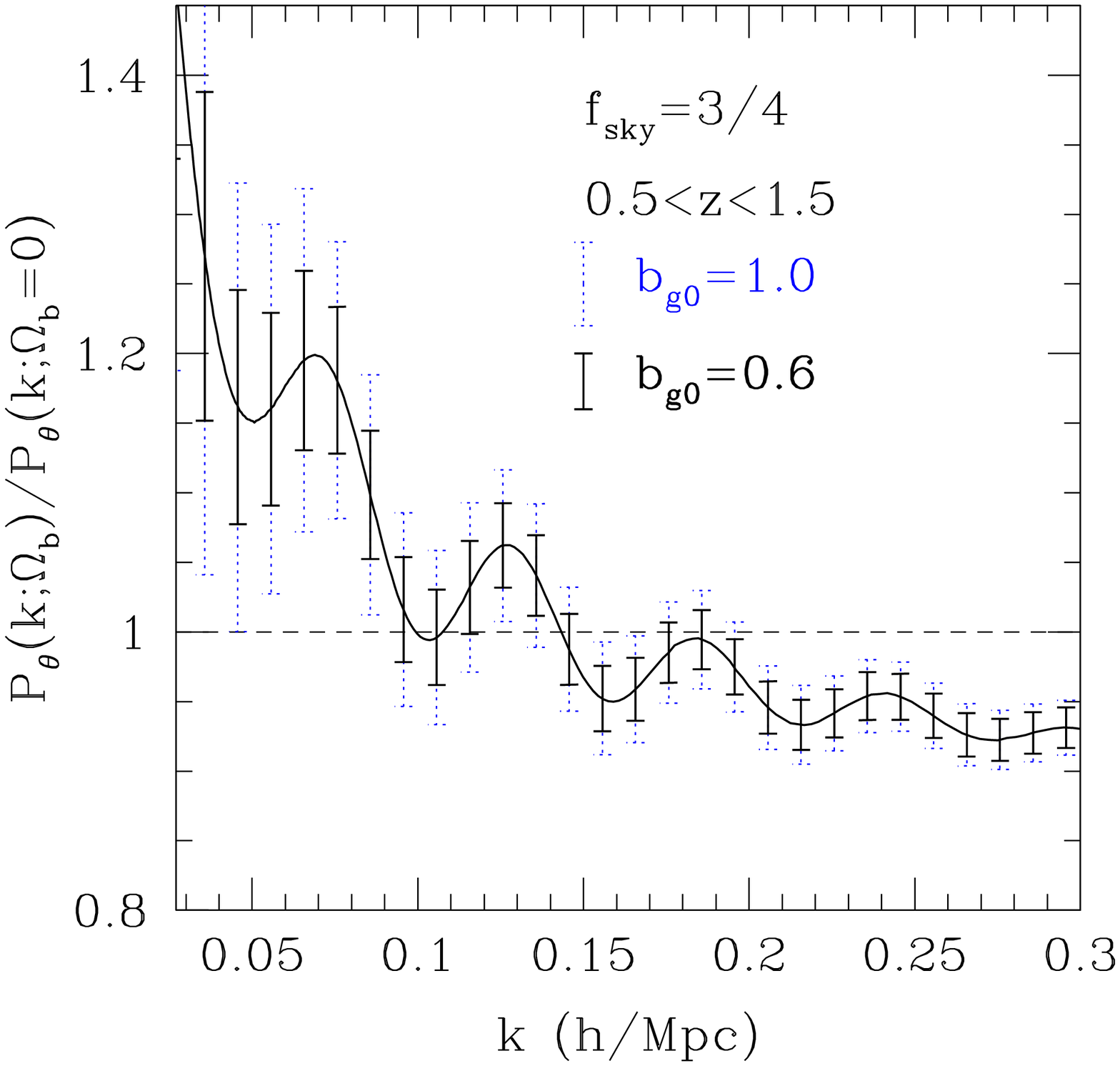}
\caption{Baryon wiggles in $P_{\theta}$ of $0.5<z<1.5$
  galaxies. The accuracy of $P_{\theta}$ reconstruction is sensitive to $b_g$. Smaller $b_g$
  causes stronger directional dependence in $P_g^s$ and weaker
  statistical fluctuations in $P_g$ and $P_{g\theta}$ and  thus improves
  $P_{\theta}$ reconstruction. SKA is able to detect numerous galaxies
  with weak clustering at $z\sim 1$. $P_\theta$ reconstructed through
  those galaxies can be sufficiently accurate for BAO measurement in
  the velocity field. \label{fig:v1}} 
\efi
%%%%%%%%%%%%%%%%%%%%%%%%%%%%%

Accuracy in $P_\theta$ reconstruction is sensitive to the galaxy bias $b_g$. A
smaller bias causes a stronger directional dependence and enables better
$P_\theta$ reconstruction.   At the first three peaks of BAO, shot
noise is in general sub-dominant 
for SKA. This allows us the luxury to select those low mass, weakly clustered
galaxies for better BAO  measurement in the velocity field. A
practical question is that, how strong are SKA galaxies
clustered. Observationally, we know that galaxies in local universe detected through 21cm
line emission are weakly clustered \citep{Meyer07}. However, we are
lack of observations on $b_g$ of neutral hydrogen mass selected galaxies other
than in the local universe. For this reason,  we just present a
rough estimation here.  

SKA galaxies are selected by their neutral hydrogen mass $M_{\rm
  HI}$. We update the calculation of lower $M_{\rm HI}$ limit in
\citet{Zhang06} with new SKA specifications. In a ten year survey,
$M_{\rm HI}$ at  $3\sigma$ selection threshold is shown in Fig.\ref{fig:bias}.
To calculate the associated bias of these galaxies, we
need to convert $M_{\rm HI}$ to  the galaxy total mass $M$. We define the
neutral  hydrogen fraction $f_{\rm HI}\equiv M_{\rm HI}/M$. The value
of $f_{\rm HI}$ is poorly known. But we are able to derive its lower
limit. The total neutral hydrogen density is 
$\Omega_{\rm HI}\simeq 3.5\times 10^{-4}$ today \citep{Zwaan05} and increases
by a factor of $\sim 5$ 
toward $z\sim 2-3$ \citep{Peroux03}. Since most baryons and dark matter  are
not in galaxies while 
most neutral hydrogen atoms are in galaxies, we expect that $f_{\rm HI}\gg
\Omega_{\rm HI}/\Omega_m$.  For several choices of $f_{\rm HI}$, we plot the
corresponding $b_g$ in Fig. \ref{fig:bias}. We adopt  the $b_g$-$M$
relation from \citet{Jing98}, whose fitting 
formulae improves over previous analytical results, especially for low
mass halos.   Many of SKA detected galaxies at $z=1$ are likely well below
$10^{11}M_{\sun}$.  These galaxies are indeed weakly clustered, with
bias as low as $b_g\sim 0.6$.  These galaxies are very likely numerous, if
the shape of neutral hydrogen mass  function does not change at low
mass end, from local universe to $z\sim 1$.  Selecting only these
galaxies then does not result in significant increase in shot noise. What we
gain is a factor of $\sim 2$ improvement in $P_{\theta}$
reconstruction and comparable improvement in BAO measurement, comparing to the case of $b_{g0}=1$ (Fig. \ref{fig:v1}).

\section{Discussions and summary}
\label{sec:discussion}
Up to now we only discussed the feasibility to perform self calibration
of galaxy bias and to measure BAO in the velocity power spectrum
reconstructed  from redshift distortion of galaxies. This technique
can be applied to any tracers of the large scale structure with
precision redshift information. This includes the pseudo 21cm
background composed of unresolved 21cm emitting galaxies at $z\sim 1$
\citep{Chang07} and the 21cm background at higher redshifts
\citep{Mao07,Wyithe07}.

It is also feasible to measure BAOs in the velocity field
reconstructed from redshift distortion. A fundamental obstacle must be
overcome is the statistical fluctuations associated with the
reconstruction technique.  Since statistical
fluctuations in $P_g$ and $P_{g\theta}$ propagate 
into $P_{\theta}$ reconstruction, 
statistical errors in  $P_\theta$ reconstructed  are
a factor of $\sim 10$ times  larger than its cosmic variance
(Fig. \ref{fig:v2}). To reach the cosmic variance limit of $P_\theta$
and improve  significantly on BAO measurements in velocity field, new
techniques of velocity measurement are needed. Kinetic Sunyaev
Zel'dovich effect of galaxy clusters is a possibility. It is also
possible to do  precision peculiar velocity
measurement through  millions of  SNe Ia with
spectroscopic redshifts at $z\sim 0.5$ \citep{Zhang07b}.  BAO measurement brings
extra scientific benefit for such surveys.

$P_{\theta}$ reconstructed through the galaxy spectroscopic surveys contain
valuable information of cosmology, besides the information of the expansion of
the universe contained in the BAO. Combining the BAO measurements through
galaxy spectroscopic surveys and $P_{\theta}$, we are able to infer the
initial fluctuations of the universe, the structure growth of the
universe and the geometry of the universe simultaneously.  This
potentially very  promising and powerful application  will be
addressed in a companion paper. $P_{\theta}$ and 
 $P_{g\theta}$, when combined with gravitational lensing,  also have
important applications in probing the nature  of gravity at
cosmological scales \citep{Zhang07,Zhang08}. Finally
we emphasize that all these  applications rely on precision modeling of redshift
 distortion. Tremendous work is required to meet the requirement of
 precision cosmology.  

\acknowledgments
{\it Acknowledgments}: 
We thank Ravi Sheth for useful conversations and Ue-Li Pen for useful
comments. This work is supported  by the one-hundred-talents  
program of the Chinese academy of science (CAS), the national science
foundation of China grant  10533030, the CAS grant
KJCX3-SYW-N2 and the 973 program grant No. 2007CB815401.

\end{document}